\documentclass{gretsi}

\usepackage[latin1]{inputenc}
\usepackage[french,english]{babel}
\usepackage{graphicx,afterpage}
\usepackage{fancybox}
\usepackage{psboxit,pstricks,rotating}
\usepackage{psfrag}
\usepackage{ulem}
\usepackage{amsmath,bbm} 
\usepackage{amsfonts} 
\usepackage{amssymb}
\usepackage{vector}
\usepackage{floatflt}
\usepackage{theorem}
\usepackage{stmaryrd}
\usepackage{euscript}
\usepackage{pifont}
\usepackage{subfigure}
\usepackage{array}
\usepackage[T1]{fontenc}
\usepackage{endnotes}

\def\R{{\mathbb{R}}}
\def\Z{{\mathbb{Z}}}

\newtheorem{theorem}{Th\'eor\^eme}

\newtheorem{proposition}{Proposition}
\newtheorem{corollary}{Corollaire}
\newtheorem{proof}{Preuve}
\newtheorem{definition}{D\'efinition}

\title{S\'eparation en composantes structures, textures et bruit d'une image, apport de l'utilisation des contourlettes.}
\auteur{\coord{J\'er\^ome}{Gilles}{1}}
\adresse{\affil{1}{DGA-CEP/EORD, 16bis rue Prieur de la C\^ote d'Or 94110 Arcueil}}
\email{jerome.gilles@etca.fr}

\frenchabstract{Dans cet article, nous proposons une am\'elioration des m\'ethodes de d\'ecomposition d'image dans le cas d'images bruit\'ees. 
Dans \cite{gilles1,aujoluvw}, les auteurs proposent de s\'eparer les structures, textures et bruit d'une image. Malheureusement, l'utilisation 
d'ondelettes s\'eparables provoque des art\'efacts. Nous proposons ici de remplacer les ondelettes par l'utilisation des contourlettes qui permettent 
de mieux approximer la g\'eom\'etrie dans les images. Pour cela, nous d\'efinissons les espaces de contourlettes et leurs normes associ\'ees. Nous 
obtenons un algorithme it\'eratif que nous appliquons sur deux images textur\'ees et bruit\'ees.}

\englishabstract{In this paper, we propose to improve image decomposition algorithms in the case of noisy images. In \cite{gilles1,aujoluvw}, the authors propose to separate structures, textures and noise from an image. Unfortunately, the use of separable wavelets shows some artefacts. In this paper, we propose to replace the wavelet transform by the contourlet transform which better approximate geometry in images. For that, we define contourlet spaces and their associated norms. Then, we get an iterative algorithm which we test on two noisy textured images.}

\begin{document}

\maketitle

\section{Introduction}
Ces derni\'eres ann\'ees, des mod\`eles permettant de d\'ecomposer une image $f$ en ses composantes structures $u$ + textures $v$ ont vu le jour. 
Le principe, propos\'e par Y.Meyer dans \cite{meyer}, consiste \`a consid\'erer les structures comme des fonctions de l'espace des fonctions \`a variations 
born\'ees $BV$ et les textures comme des fonctions oscillantes appartenant \`a un espace not\'e $G$ proche du dual de $BV$. Le mod\`ele consiste alors \`a 
minimiser la fonctionnelle
\begin{equation}\label{eqn:eq1}
F^{YM}(u,v)=\|u\|_{BV}+\lambda\|v\|_G.
\end{equation}
Un algorithme num\'erique efficace it\'eratif bas\'e sur un projecteur non lin\'eaire $P_{G_{\lambda}}$ a \'et\'e propos\'e par J.F Aujol dans \cite{aujol2} 
moyennant une modification de la fonctionnelle (mais donnant les m\^emes minimiseurs que pour (\ref{eqn:eq1})):
\begin{equation}\label{eq:aujolg}
F_{\lambda,\mu}^{AU}(u,v)=J(u)+J^*\left(\frac{v}{\mu}\right)+(2\lambda)^{-1}\|f-u-v\|_{L^2}^2
\end{equation}
o\`u $J^*(v)$ est la fonction caract\'eristique sur $G_1$ o\`u $G_{\mu}=\{v\in G / \|v\|_G\leqslant \mu\}$ et $J(u)=\|u\|_{BV}$.

Ce mod\`ele montre ses limites d\`es que les images d'entr\'ee sont bruit\'ees. En effet, le bruit peut \^etre consid\'er\'e comme un signal tr\`es oscillant 
et sera donc captur\'e dans la composante $v$. Il est alors n\'ecessaire d'\'etendre ce mod\`ele \`a un mod\`ele \`a trois composantes: structures $u$ + 
textures $v$ + bruit $w$. Dans \cite{gilles1}, nous avons propos\'e une premi\`ere approche permettant de r\'ealiser cette d\'ecomposition en jouant d'une part 
sur les bornes sup\'erieures de la norme dans l'espace $G$ pour les composantes $v$ et $w$ et d'autre part en donnant un comportement localement adaptatif \`a 
l'algorithme. Nous avons compar\'e ce mod\`ele \`a celui propos\'e par Aujol et al. \cite{aujol1} utilisant un seuillage des coefficients d'une d\'ecomposition 
en ondelettes pour effectuer le d\'ebruitage. Les deux m\'ethodes donnent de bon r\'esultats, le seuillage des ondelettes montrant une meilleure performance de 
d\'ebruitage mais ab\^imant les bords des structures. Ce ph\'enom\`ene est d\`u au fait que la transform\'ee en ondelette 2D se base sur deux directions. 
Aussi dans cette communication, nous proposons de remplacer l'utilisation des ondelettes par celle des contourlettes \cite{do}. En effet, cette repr\'esentation 
permet de mieux tenir compte de la g\'eom\'etrie des bords pr\'esents dans les images. Nous d\'efinissons la notion d'espace des contourlettes et montrons que 
le seuillage des coefficients revient \`a faire la projection sur cet espace.

\section{S\'eparation structures, textures et bruit}
Commen\c{c}ons par rappeler les mod\`eles propos\'es dans \cite{gilles1} et \cite{aujol1} permettant d'effectuer cette d\'ecomposition en trois composantes. 
Le premier consid\`ere que les textures et le bruit sont des fonctions oscillantes \`a ceci pr\`es que le bruit est vu comme beaucoup plus oscillant. Suivant 
les propri\'et\'es de la norme sur l'espace $G$ (cette norme est d'autant plus faible que le signal est oscillant), nous consid\'erons que $v\in G_{\mu_1}$ et 
$w\in G_{\mu_2}$ avec $\mu_2 \ll \mu_1$. De plus, nous proposons d'utiliser une \og{}carte\fg{} $\nu$ des zones o\`u sont pr\'esentes des textures afin 
d'accentuer ou non le pouvoir de d\'ebruitage (voir \cite{gilles1} pour plus de d\'etails). La fonctionnelle \`a minimiser est:
\begin{align}
F_{\lambda ,\mu_1 ,\mu_2}^{JG}(u,v,w)= J(u)&+J^*\left(\frac{v}{\mu_1}\right)+J^*\left(\frac{w}{\mu_2}\right)\\ \notag
&+(2\lambda)^{-1}\|f-u-\nu_1 v-\nu_2 w\|_{L^2}^2.
\end{align}
Moyennant une l\'eg\`ere modification, nous pouvons utiliser les projecteurs non lin\'eaires utilis\'es dans le cas \`a deux composantes. Tous les d\'etails 
th\'eoriques et num\'eriques sont disponibles dans \cite{gilles1}.

Dans \cite{aujol1}, les auteurs proposent un mod\`ele assez similaire mais consid\'erant le bruit comme appartenant \`a l'espace des distributions. Cette espace 
\'etant formul\'e via l'espace de Besov $E=\dot{B}_{-1,\infty}^{\infty}$. La fonctionnelle propos\'ee est alors:
\begin{align}\label{eqn:aujoluvw}
F_{\lambda,\mu,\delta}^{AC}(u,v,w)= J(u)&+J^*\left(\frac{v}{\mu}\right)+B^*\left(\frac{w}{\delta}\right)\\ \notag
&+(2\lambda)^{-1}\|f-u-v-w\|_{L^2}^2,
\end{align}
o\`u $B^*(w)$ est la fonction caract\'eristique sur l'espace $E_1=\{w\in E / \|w\|_E \leqslant 1\}$. Les auteurs montrent que la projection sur cet espace 
correspond \`a effectuer un seuillage doux des coefficients d'une d\'ecomposition en ondelettes. Ce mod\`ele donne un meilleur d\'ebruitage mais ab\^ime les 
bords des structures, d\^u au fait que la transform\'ee 2D n'utilise que des filtres s\'eparables horizontaux et verticaux. Nous proposons donc de remplacer 
l'utilisation de ces ondelettes par une repr\'esentation mieux adapt\'ee au traitement des images et plus particuli\`erement \`a la g\'eom\'etrie pr\'esente. 
La r\'epr\'esentation retenue est celle des contourlet\-tes \cite{do}. 

\section{S\'eparation de composantes bas\'ee sur les contourlettes}
Nous commen\c{c}ons par rappeler un r\'esultat d\'emontr\'e par Do et Vetterli \cite{do}:

\begin{theorem}
Soit $j$ l'\'echelle, $n$ la position, $\left\{l_j\right\}_{j\leqslant j_0}$ l'en\-semble des nombres de directions souhait\'ees pour chaque \'echel\-le $j$. 
Alors, la famille
\begin{equation}
\left\{\phi_{j_0,n}(t);\rho^{(l_j)}_{j,k,n}(t)\right\}_{j\leqslant j_0, \; 0\leqslant k\leqslant 2^{l_j}-1, \;n\in\Z^2}
\end{equation}
est une trame ajust\'ee de $L_2(\R^2)$.
\end{theorem}

Tous les d\'etails sur la construction des fonctions $\phi_{j_0,n}(t)$ et $\rho^{(l_j)}_{j,k,n}(t)$ sont disponibles dans \cite{do,do4,do5}. Ceci implique 
donc que l'on peut d\'ecomposer une fonction de $L_2(\R^2)$ de la mani\`ere suivante:

\begin{corollary}\label{cor:contourlet}
\begin{equation}
f(t)=\sum_n \alpha_n \phi_{j_0,n}(t)+\sum_{j\leqslant j_0}\sum_{k=0}^{2^{l_j}-1}\sum_n \beta_{j,k,n}\rho^{(l_j)}_{j,k,n}(t)
\end{equation}
ou
\begin{equation}
f(t)=\sum_{j\in \Z}\sum_{k=0}^{2^{l_j}-1}\sum_n \beta_{j,k,n}\rho^{(l_j)}_{j,k,n}(t).
\end{equation}
\end{corollary}

O\`u $\alpha_n=\langle f|\phi_{j_0,n}\rangle$ et $\beta_{j,k,n}=\langle f|\rho_{j,k,n}^{(l_j)}\rangle$ sont les coefficents de la transform\'ee en 
contourlettes. Nous pouvons alors d\'efinir, sur le mod\`ee des espaces de Besov, les espaces de contourlettes $Co_{p,q}^{s}$ et leurs versions homog\`enes 
$\dot{Co}_{p,q}^{s}$ ainsi que leurs normes associ\'ees:
\begin{definition}
Soit $s\geqslant 0$ et $p,q>0$, si $f\in Co_{p,q}^s$ alors
\begin{align} \label{eq:defctspq}
&\|f\|_{Co_{p,q}^s}=\left[\sum_n |\alpha_{j_0,n}|^p \right]^{1/p} \notag\\ 
&+\left\{\sum_{j\leqslant j_0}2^{j\left(\frac{d}{2}-\frac{1}{p}+s\right)q}\left[\sum_{k=0}^{2^{l_j}-1}\sum_n 2^{j\frac{p}{2}}|\beta_{j,k,n}|^p\right]^{q/p}\right\}^{1/q}
\end{align}
ou dans le cas des espaces homog\`enes
\begin{align} \label{eq:defctspq2}
&\|f\|_{\dot{Co}_{p,q}^s}=\notag\\
&\left\{\sum_{j\in \Z}2^{j\left(\frac{d}{2}-\frac{1}{p}+s\right)q}\left[\sum_{k=0}^{2^{l_j}-1}\sum_n 2^{j\frac{p}{2}}|\beta_{j,k,n}|^p\right]^{q/p}\right\}^{1/q}
\end{align}
\end{definition}

Il est rapide de voir que l'espace $\dot{Co}_{-1,\infty}^{\infty}$, comme dans le cas des espaces de Besov, contient les distributions et peut donc \^etre 
utilis\'e pour mod\'eliser le bruit dans l'image. La fonctionnelle prenant en compte cet espace permettant la d\'ecomposition est alors:

\begin{align}\label{eqn:uvwct}
F_{\lambda,\mu,\delta}^{Co}(u,v,w)=J(u)&+J^*\left(\frac{v}{\mu}\right)+J_{Co}^*\left(\frac{w}{\delta}\right)\\ \notag
&+(2\lambda)^{-1}\|f-u-v-w\|_{L^2}^2
\end{align}

Avec $J_{Co}^*$ la fonction caract\'eristique sur $Co_1$ o\`u 
$Co_{\delta}=\left\{f\in \dot{Co}_{-1,\infty}^{\infty} / \|f\|_{\dot{Co}_{-1,\infty}^{\infty}}\leqslant \delta \right\}$. 
Les minimiseurs de (\ref{eqn:uvwct}) sont donn\'es par la proposition suivante.

\begin{proposition}\label{pro:uvwjg3}
Soit $u\in BV$, $v\in G_{\mu}$, $w\in Co_{\delta}$ respectivement les composantes structures, textures et bruit d\'ecoulant de la d\'ecomposition d'image. 
Alors, les minimiseurs
\begin{equation}
(\hat{u},\hat{v},\hat{w})=\underset{(u,v,w)\in BV\times G_{\mu}\times Co_{\delta}}{\arg} \inf F_{\lambda,\mu,\delta}^{Co}(u,v,w)
\end{equation}
sont donn\'es par
\begin{align*}
\hat{u}&=f-\hat{v}-\hat{w}-P_{G_{\lambda}}(f-\hat{v}-\hat{w}) \\
\hat{v}&=P_{G_{\mu}}\left(f-\hat{u}-\hat{w}\right) \\
\hat{w}&=f-\hat{u}-\hat{v}-CST\left(f-\hat{u}-\hat{v};2\delta\right)
\end{align*}
o\`u $P_{G_{\lambda}}$ est le projecteur non-lin\'eaire de Chambolle et $CST$ $(f-u-v,2\delta)$ est l'op\'erateur de seuillage doux, avec un seuil de $2\delta$, 
des coefficients de la transform\'ee en contourlettes de $f-u-v$.\\
\end{proposition}

\begin{proof}
Les composantes $\hat{u}$ et $\hat{v}$ sont obtenues par le m\^eme argument utilis\'e dans les d\'emonstrations des travaux de Aujol et Chambolle 
\cite{aujoluvw} ou dans les travaux de Gilles \cite{gilles1} et ne seront pas red\'emontr\'es ici. Le point particulier concerne l'expression de $\hat{w}$ en 
fonction du seuillage doux des coefficients de la transform\'ee en contourlettes. Supposons que l'on cherche \`a minimiser $F_{\lambda,\mu,\delta}^{Co}(u,v,w)$ 
par rapport \`a $w$, ceci est equivalent \`a chercher $w$ solution de (on pose $g=f-u-v$)
\begin{equation}
\hat{w}=\underset{w\in Co_{\delta}}{\arg}\min \left\lbrace \|g-w\|_{L^2}^2\right\rbrace.
\end{equation}
Nous pouvons alors utiliser une formulation duale: $\hat{w}=g-\hat{h}$ telle que
\begin{equation}\label{eq:dualct}
\hat{h}=\underset{h\in \dot{Co}_{1,1}^1}{\arg}\min \left\lbrace 2\delta\|h\|_{\dot{Co}_{1,1}^1}+\|g-h\|_{L^2}^2\right\rbrace. 
\end{equation} 
Nous pouvons utiliser la m\^eme approche que Chambolle et al. dans \cite{chambolle2}. 

Soit $(c_{j,k,n})_{j\in\Z , 0\leqslant k\leqslant 2^{(l_j)},n\in\Z^2}$ et $(d_{j,k,n})_{j\in\Z , 0\leqslant k\leqslant 2^{(l_j)},n\in\Z^2}$ les coefficients 
respectivement issus de la transform\'ee en contourlettes de $g$ et $h$. Comme les contourlettes forment une trame ajust\'ee, de borne $1$, nous avons (on note 
$\Omega =\Z \times \llbracket 0,2^{(l_j)} \rrbracket\times \Z^2$)
\begin{equation}
\|g\|_{L_2}^2=\sum_{(j,k,n)\in\Omega}|c_{j,k,n}|^2.
\end{equation}
 
Alors (\ref{eq:dualct}) consiste \`a trouver les $d_{j,k,n}$ minimisant
\begin{equation}
\sum_{(j,k,n)\in \Omega}|c_{j,k,n}-d_{j,k,n}|^2+2\delta\sum_{(j,k,n)\in \Omega}|d_{j,k,n}|
\end{equation}
ce qui est equivalent \`a minimiser
\begin{equation}
|c_{j,k,n}-d_{j,k,n}|^2+2\delta|d_{j,k,n}|.
\end{equation}
Or, dans \cite{chambolle2}, les auteurs montrent que la solution \`a ce type de probl\`eme est un seuillage doux des coefficients $(c_{j,k,n})$ avec un seuil 
de $2\delta$.

Alors $\hat{h}=CST(g,2\delta)$, ce qui, par dualit\'e, implique que $\hat{w}=g-CST(g,2\delta)$. Nous concluons donc que
\begin{equation}
\hat{w}=f-\hat{u}-\hat{v}-CST(f-\hat{u}-\hat{v},2\delta)
\end{equation}
ce qui termine la preuve.
\end{proof}

%==================
L'algorithme num\'erique correspondant consiste donc en celui de Aujol et al. dans lequel nous rempla\c{c}ons l'utilisa\-tion des ondelettes par celle des 
contourlettes.

\begin{center}
\fbox{\parbox[h]{0.48\textwidth}{
\begin{enumerate}
\item initialisation: $u_0=v_0=w_0=0$,
\item calcul de $w_{n+1}=f-u_n-v_n-CST(f-u_n-v_n,2\delta)$,
\item calcul de $v_{n+1}=P_{G_{\mu}}(f-u_n-w_{n+1})$,
\item calcul de $u_{n+1}=f-v_{n+1}-w_{n+1}-P_{G_{\lambda}}(f-v_{n+1}-w_{n+1})$,
\item si $\max\{|u_{n+1}-u_n|,|v_{n+1}-v_n|,|w_{n+1}-w_n|\}\leqslant\epsilon$ ou si $N_{step}$ it\'erations ont \'et\'e faites alors on stoppe l'algorithme, 
sinon on retourne \`a l'\'etape 2.
\end{enumerate}}}
\end{center}

\section{R\'esultats}
\begin{figure}[!ht]
\centering
\begin{tabular}{cc}
\includegraphics[width=0.22\textwidth]{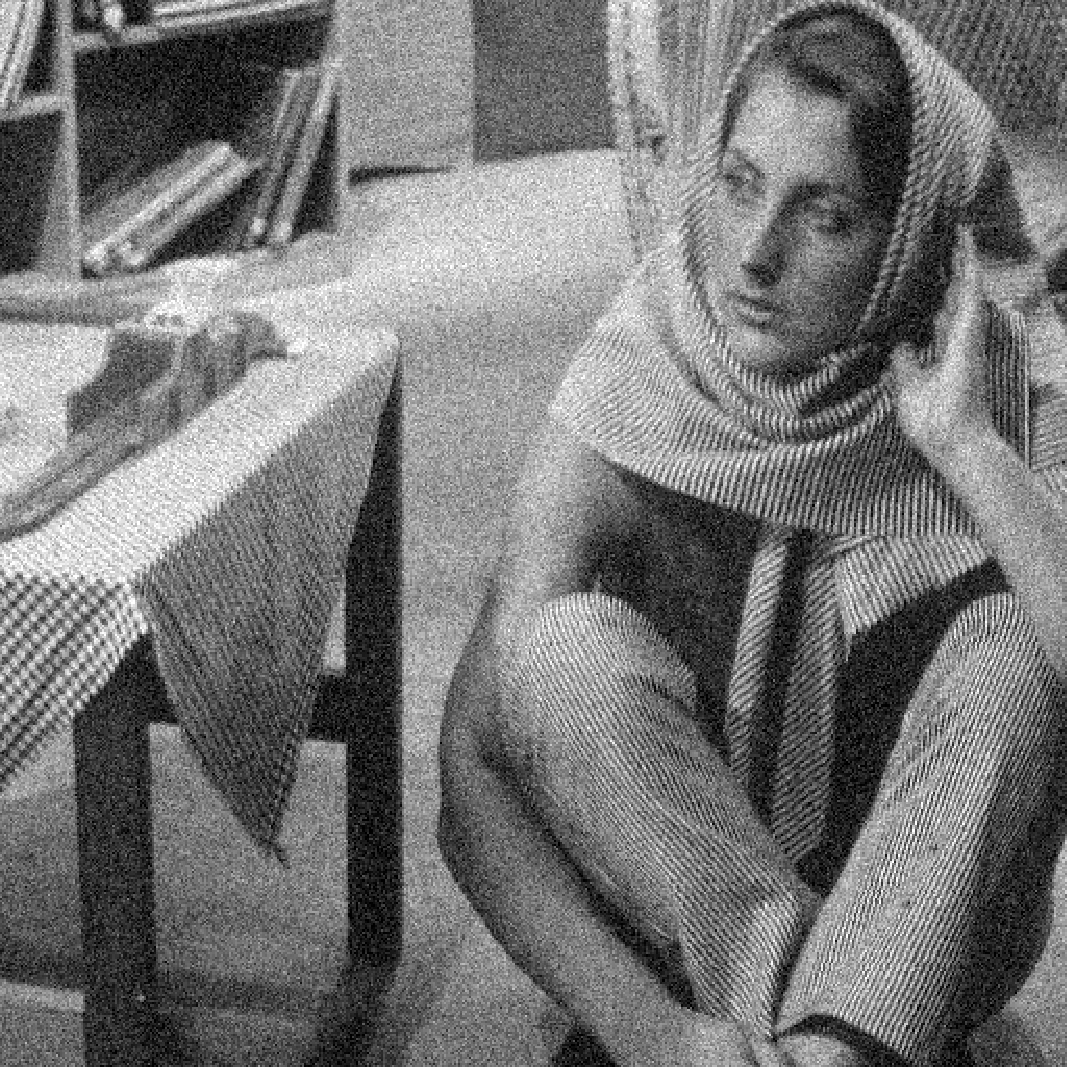} & \includegraphics[width=0.22\textwidth]{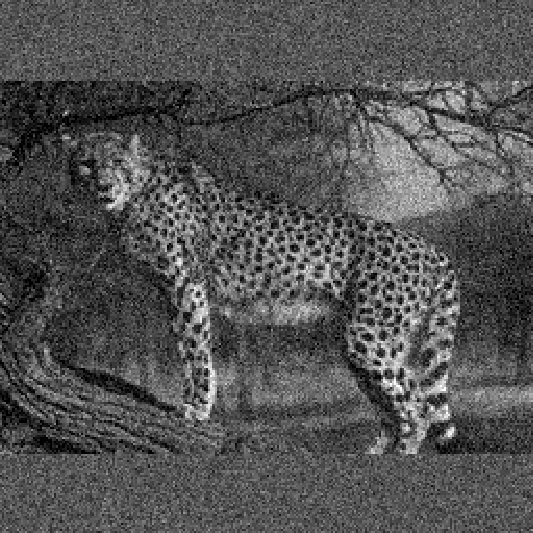} 
\end{tabular}
\caption{Images de test Barbara et L\'eopard corrompues par un bruit gaussien ($\sigma=20$).}
\label{fig:testimage}
\end{figure}

La figure \ref{fig:testimage} montre les images de test utilis\'ees. Nous leur avons ajout\'e un bruit de type gaussien additif d'une variance $\sigma=20$ sur 
chaque image.

\begin{figure}[!ht]
\centering
%\begin{tabular}{m{5cm}m{5cm}}
\begin{tabular}{cc}
\includegraphics[width=0.22\textwidth]{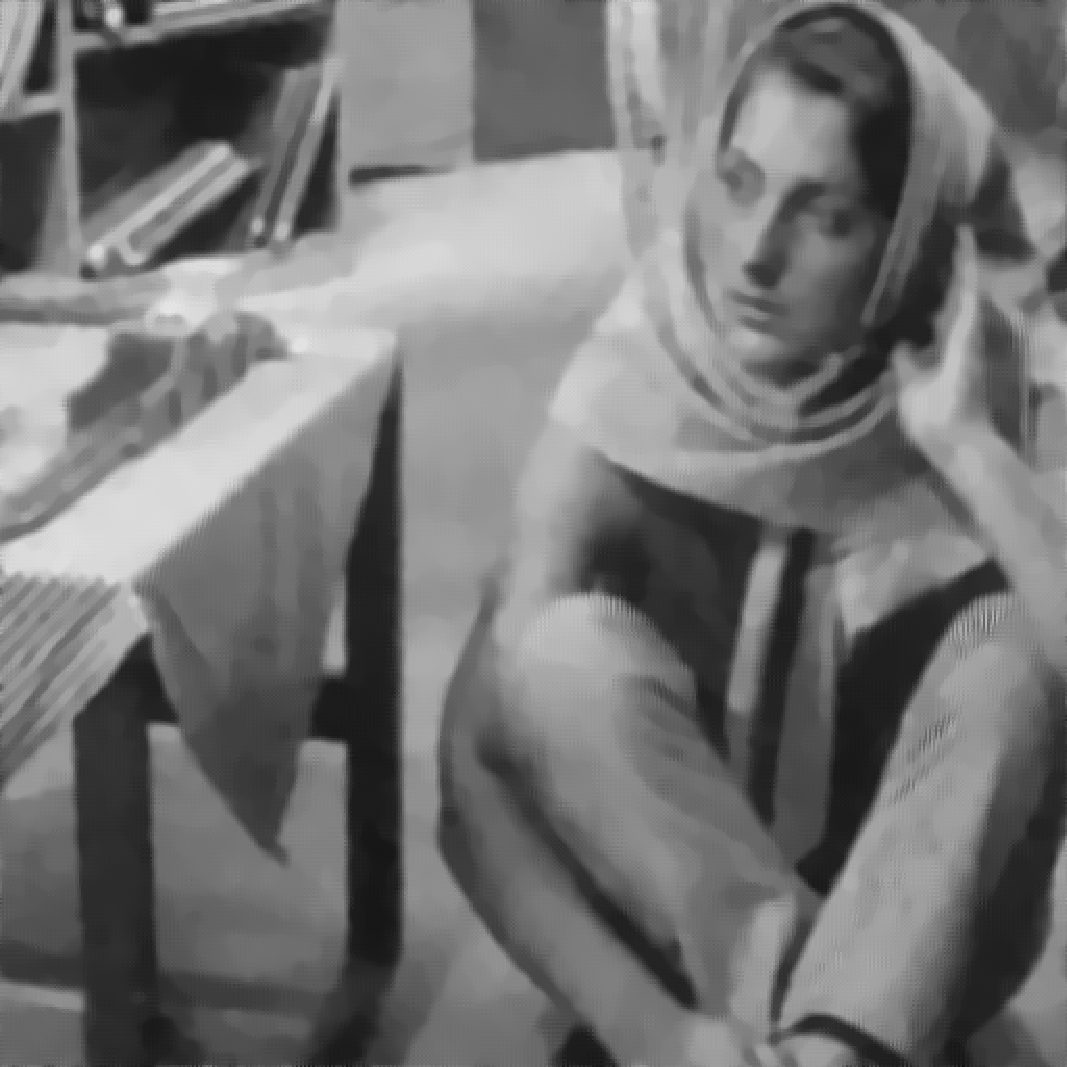} &  \includegraphics[width=0.22\textwidth]{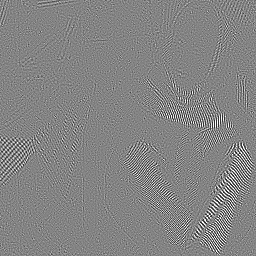} \\
Structures & Textures \\
\includegraphics[width=0.22\textwidth]{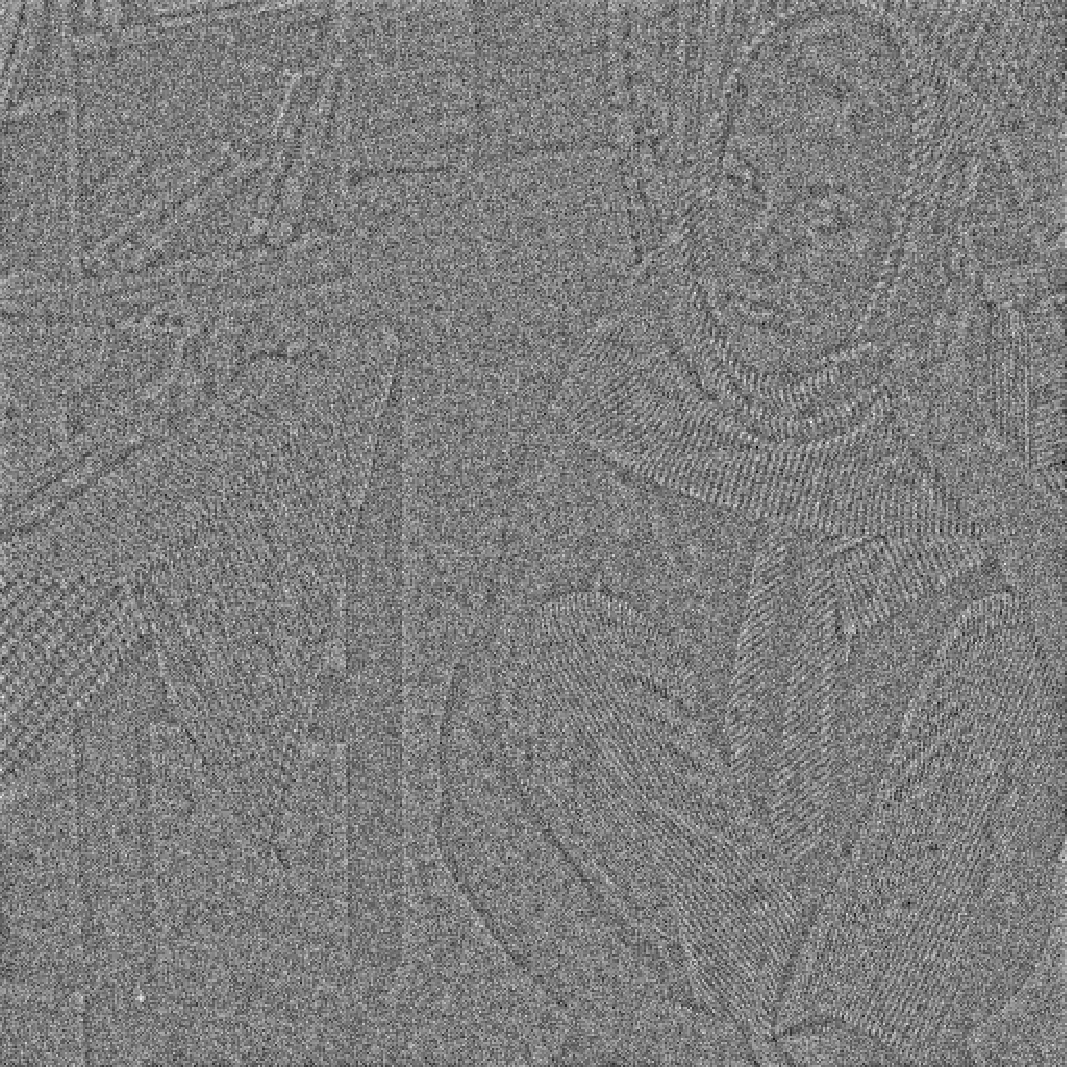} & \\
Bruit &
\end{tabular}
\caption{Composantes structures $+$ textures $+$ bruit issus de la d\'ecomposition de Barbara bruit\'ee.}
\label{fig:uvwcobarb}
\end{figure}

\begin{figure}[!ht]
\centering
%\begin{tabular}{m{5cm}m{5cm}}
\begin{tabular}{cc}
\includegraphics[width=0.22\textwidth]{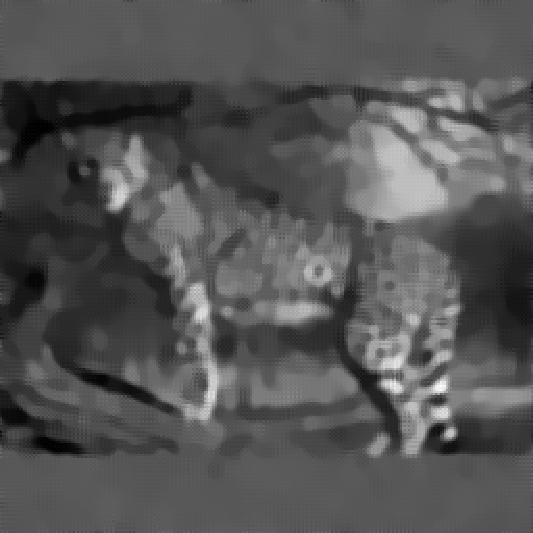} &  \includegraphics[width=0.22\textwidth]{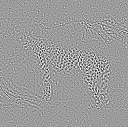} \\
Structures & Textures \\
\includegraphics[width=0.22\textwidth]{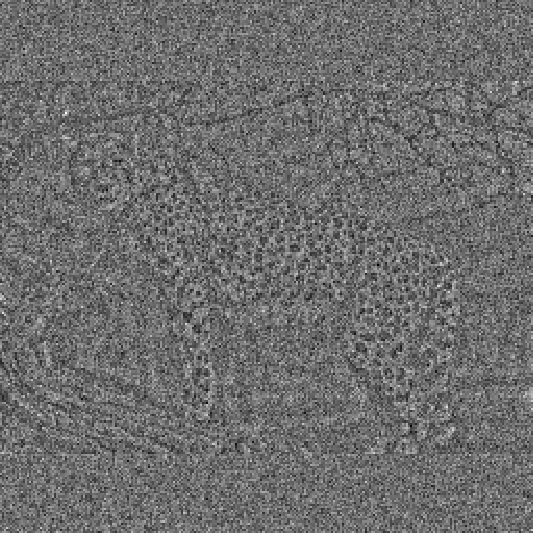} & \\
Bruit &
\end{tabular}
\caption{Composantes structures $+$ textures $+$ bruit issus de la d\'ecomposition du L\'eopard bruit\'e.}
\label{fig:uvwcoleo}
\end{figure}

Les r\'esultats de la d\'ecomposition obtenue sur chaque image sont donn\'es respectivement sur les figures \ref{fig:uvwcobarb} et \ref{fig:uvwcoleo}. Le bruit 
est effectivement bien s\'epar\'e du reste des autres composantes. Nous pouvons toutefois remarquer, comme pour les autres algorithmes cit\'es pr\'ec\'edemment, 
que le choix des param\^etres influe sur la quantit\'e de r\'esidu de bruit dans les textures et la quantit\'e de r\'esidu de textures dans le bruit. Par 
ailleurs, comme attendu, les structures sont de meilleure qualit\'e du fait de l'utilisation des contourlettes.

\section{Conclusion}
Dans cet article, nous avons pr\'esent\'e une extension du mod\`ele de d\'ecomposition d'image bas\'e sur le seuillage de coefficients d'ondelette propos\'e par 
Aujol et al. Pour cela, nous avons choisi d'utiliser les contourlettes. Celles-ci permettent en effet de tenir compte de la g\'eom\'etrie dans les images et ont 
un meilleur pouvoir d'approximation que les ondelettes. Nous d\'efinissons alors les espaces de contourlettes ainsi que leurs normes associ\'ees; et montrons 
que la projection sur ces espaces correspond, comme dans le cas des ondelettes, \`a un seuillage doux des coefficients de la transform\'ee.

L'algorithme final permettant d'obtenir la d\'ecomposition en trois composantes revient au m\^eme algorithme que dans le cas des ondelettes \`a la diff\'erence 
pr\`es que le seuillage des coefficients d'ondelette est rempla\c{c}\'e par le seuillage des coefficients de contourlettes.

Les r\'esultats obtenus montrent effectivement un gain, d'une part sur la qualit\'e des structures extraites, d'autre part sur le pouvoir de d\'ebruitage (et 
donc le pouvoir de s\'eparabilit\'e du bruit et des textures).

Il serait int\'eressant, dans le futur, d'utiliser de nouvelles repr\'esentations comme les bandlettes ou encore les repr\'esentations \og{}creuses\fg{} qui 
permettrait certainement de gagner encore sur la qualit\'e des composantes extraites.

\end{document}